\documentclass[runningheads]{llncs}
\usepackage[T1]{fontenc}
\usepackage{graphicx}
\usepackage{amsmath,amssymb}
\usepackage[misc]{ifsym}

\usepackage[marginal]{footmisc}

\begin{document}
\title{SRE-CNN: A Spatiotemporal Rotation-Equivariant CNN for Cardiac Cine MR Imaging }
\titlerunning{SRE-CNN: A Spatiotemporal Rotation-Equivariant CNN}
\author{Yuliang Zhu\inst{1,2} \and 
Jing Cheng\inst{2} \and
Zhuo-Xu Cui\inst{2} \and
Jianfeng Ren\inst{1} \and 
\\
Chengbo Wang\inst{1} \and
Dong Liang\inst{2,3}$^{(\textrm{\Letter})}$ }
%
% First Author 1 {Zhu, Yuliang},
%
% First Author 2 {Cheng, Jing},
% Second Author 3 {Cui, Zhuo-Xu},
% Third Author 4 {Ren, Jianfeng},
% Fourth Author 5 {Wang, Chengbo},
% Corresponding Author 6 {Liang, Dong}.

\authorrunning{Y. Zhu et al.}
\institute{University of Nottingham Ningbo China, Ningbo, China 
\and
Shenzhen Institutes of Advanced Technology, Chinese Academy of Sciences, Shenzhen, China
\and
Key Laboratory of Biomedical Imaging Science and System, Chinese Academy of
Sciences, Shenzhen, China
\\
\email{dong.liang@siat.ac.cn} 
}
\maketitle              % typeset the header of the contribution

\footnote{Y. Zhu and J. Cheng—Contributed equally to this work.}

\begin{abstract}
Dynamic MR images possess various transformation symmetries, including the rotation symmetry of local features within the image and along the temporal dimension. Utilizing these symmetries as prior knowledge can facilitate dynamic MR imaging with high spatiotemporal resolution. Equivariant CNN is an effective tool to leverage the symmetry priors. However, current equivariant CNN methods fail to fully exploit these symmetry priors in dynamic MR imaging. In this work, we propose a novel framework of Spatiotemporal Rotation-Equivariant CNN (SRE-CNN), spanning from the underlying high-precision filter design to the construction of the temporal-equivariant convolutional module and imaging model, to fully harness the rotation symmetries inherent in dynamic MR images. The temporal-equivariant convolutional module enables exploitation the rotation symmetries in both spatial and temporal dimensions, while the high-precision convolutional filter, based on parametrization strategy, enhances the utilization of rotation symmetry of local features to improve the reconstruction of detailed anatomical structures. Experiments conducted on highly undersampled dynamic cardiac cine data (up to 20X) have demonstrated the superior performance of our proposed approach, both quantitatively and qualitatively.

\end{abstract}

\section{Introduction}
Cardiac cine magnetic resonance imaging (MRI) plays a vital role in evidence-based diagnostic of cardiovascular disease and is widely regarded as the gold-standard for assessment of heart morphology and function \cite{ref16,ref26}. Dynamic cardiac cine imaging with high spatiotemporal resolution is highly demanded for diagnosis. However, the long scan time, due to the inherent physical limitations of MRI, makes it difficult to achieve both high spatial and temporal resolution \cite{ref12}. Moreover, long scan times may induce patient discomfort and elevate the risk of image artifacts stemming from patient motion \cite{ref10}. Therefore, fast imaging to improve spatiotemporal resolution and short scan time is highly demanded for dynamic cardiac cine imaging.

Undersampling in k-space is one of the most important strategies to accelerate MR imaging \cite{ref14}. In recent years, deep neural networks, particularly convolutional neural network (CNN), have shown great success in reconstructing images from highly undersampled k-space data \cite{ref13,ref21}. One of the key factors contributing to this success is the inherent translation equivariance of regular CNNs \cite{ref7,ref9}. While apart from translation symmetry, there are also many other transformation symmetries intrinsically existing within MR images [1], such as rotation, scaling and reflection symmetries, which regular CNNs cannot directly exploit \cite{ref3}. Equivariant CNN is an effective tool to leverage these transformation symmetries and has been applied in MR image reconstruction tasks \cite{ref4,ref6,ref8}. Gunel et al. \cite{ref8} designed the scale-equivariant CNNs to improve data efficiency and robustness to drifts in scale of the MR images. Additionally, the work by Chen et al. \cite{ref4} demonstrated that rotational-equivariant CNNs can enhance the generalization and provide noise robustness in unsupervised MRI reconstruction. 

However, current equivariant CNNs in MR reconstruction only focus on the transformation symmetry of the entire image or objects, ignoring the local features in image, which refer to the detailed anatomical structures essential for diagnose. In addition, for dynamic cardiac cine imaging, the symmetry along temporal dimension has not been exploited. As shown in Fig. 1, there are numerous similar anatomical structures (e.g. the edge of myocardium) under different orientations in one image frame, and the rotation symmetry of the anatomical structures is also exhibited along the temporal dimension. Both the spatial and temporal symmetries can be leveraged to improve the utilization of symmetry priors and may facilitate the reconstruction. Nevertheless, the current 2D equivariant CNN cannot be applied to 2D+t (2D spatial plus temporal axis) dynamic data directly, as the convolution along temporal dimension would destroy the equivariance of the transformation symmetry inbuilt in the 2D equivariant CNN.

\begin{figure}[t]
\centering
\includegraphics[width=0.8\textwidth]{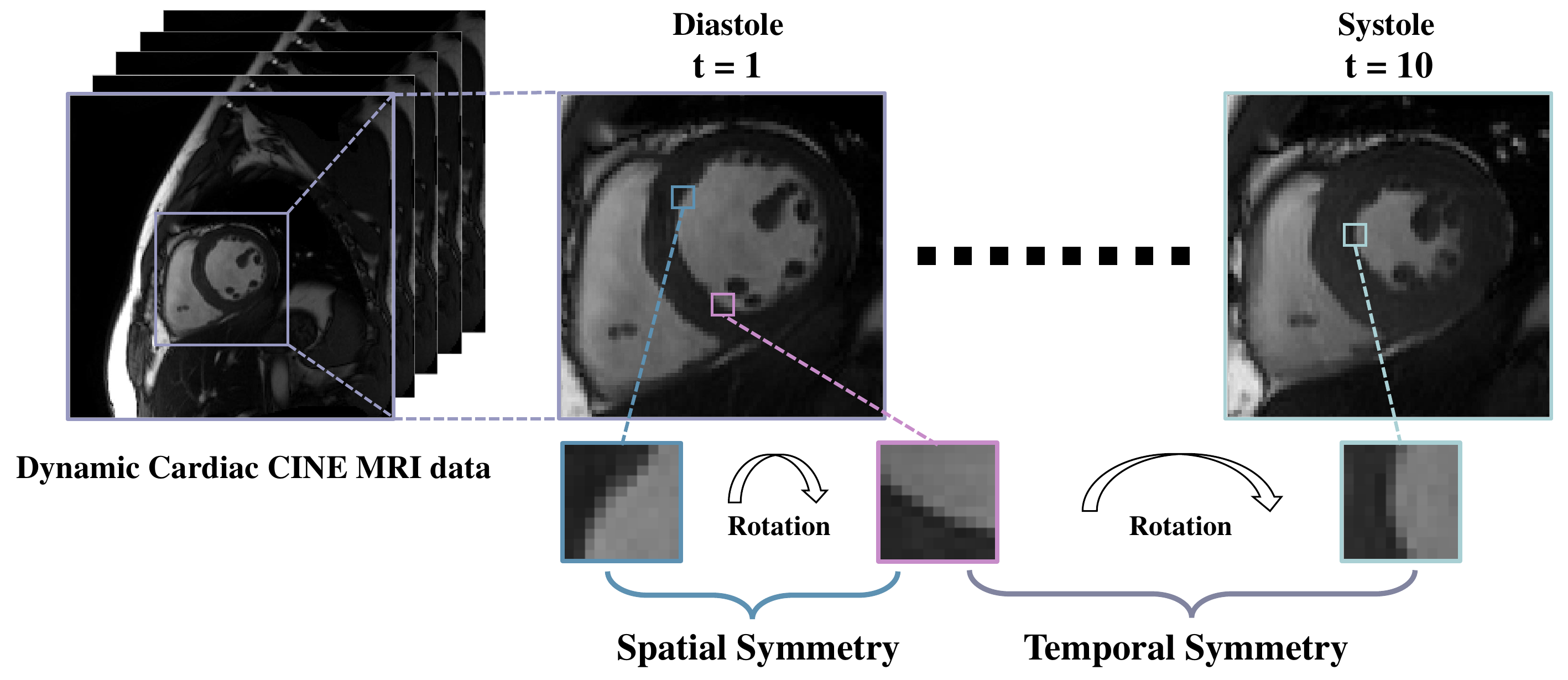}
\caption{ An example of 2D+t dynamic cardiac cine MRI data. Similar anatomical structures under different orientations are zoomed in and marked by colored bounding box.} \label{fig1}
\end{figure}

To address the issues, we introduce a Spatiotemporal Rotation-Equivariant CNN (SRE-CNN) to take full advantage of the rotation symmetry in both spatial and temporal dimensions inherent in dynamic MR cine data. Specifically, a novel temporal-equivariant convolutional module is designed to connect two spatial equivariant convolutional layers, ensuring the global equivariance of the rotation symmetry throughout the 2D+t CNNs. Furthermore, a high-precision filter parametrization strategy based on 1D and 2D Fourier series expansion is used to improve the representation accuracy of the filters, thus enhancing the performance in anatomical detail preserving. To the best of our knowledge, the proposed SRE-CNN enables the first equivariant CNN model for dynamic MRI reconstruction. The contributions of this work can be mainly summarized as follows: (1) We explore the application of equivariant CNNs to dynamic MR imaging. Specifically, the rotation symmetry encoded in the network architecture serves as an additional imaging prior in the unrolling dynamic MR reconstruction method. (2) We propose a temporal-equivariant convolutional module to preserve the global equivariance of the rotation symmetry throughout the whole network. This module enables the network to effectively exploit rotation symmetry in both spatial and temporal dimensions of 2D+t MR dynamic data. (3) We introduce a high-precision filter parametrization strategy based on 1D and 2D Fourier series expansion to improve the accuracy of convolutional filter representation, providing the possibility for reconstruction of detailed features in dynamic MR imaging. (4) The proposed method was trained and tested on in-house acquired cardiac CINE MRI data of 29 subjects and compared to state-of-the-art(SOTA) reconstruction methods. The conducted experiments demonstrate that the proposed SRE-CNN outperforms competing methods with superior qualitative and quantitative results, especially in high acceleration situations.

\section{Method}
\subsection{Accelerated MRI Reconstruction Model}
In our proposed approach, the accelerated MRI reconstruction model is based on an unrolled neural network, which unrolls the existing iterative reconstruction algorithms to deep networks \cite{ref5}. The regularized reconstruction can be formulated as a nonlinear inverse problem of the form: 
\begin{equation}
\hat{x}=arg \min_x{\left \| Ax-y \right \|_{2}^{2}+\lambda R\left (x \right )} 
\end{equation}
where $y$ is the k-space measurement, $x$ is the image to reconstructed, $A$ is the forward operator during acquisition, and the regularization term $R$ is associated with regularization strength $\lambda$. We employ the Proximal Gradient Descent (PGD) method to iteratively solve the optimization problem presented in Equation (1) by alternating between two updates: a data consistency update $z^{\left ( k \right ) } = x^{\left ( k \right ) } - \eta_k \nabla_x \left \| Ax^{\left ( k \right ) } - y \right \| _{2}^{2}$, where $\eta_k$ is the PGD step size, which is followed by a proximal update $x^{ ( k+1  )} = pro x_R(z^{(k)})$ at iteration $k$, where $pro x_R (\cdot)$ is the proximal operator of $\lambda R$ \cite{ref11}. In a deep-learning-based data-driven reconstruction, the proximal operator is replaced by a deep neural network: $x^{ ( k+1  )} = N_{\theta^{k}}(z^{(k)})$, where $N$ is a CNN whose parameters $\theta^{k}$ are learned uniquely for each iteration. The architecture of the unrolled reconstruction model is illustrated in Fig. 2(a). The data consistency and update steps are based on DL-ESPIRiT \cite{ref17}. The network takes a zero-filled reconstruction of a cardiac cine slice with its corresponding ESPIRiT maps as the input and is trained to reconstruct images which are close to fully sampled images. 

\subsection{Spatiotemporal Rotation-Equivariant CNN Design}
Equivariance can be formulated as follows \cite{ref24}. Let $f$ be a convolution mapping, $G$ is a group of transformations. $f$ is equivariant w.r.t. the action of $G$, if for any $g\in G$,
\begin{equation}
f\left ( \pi _{g}^{X} \left( x \right ) \right ) =\pi _{g}^{Y}\left (f\left ( x \right ) \right ) ,g\in G,x\in X
\end{equation}
where $x$ can be any input feature map, and $\pi _{g}^{(\cdot)}$ denotes a group action in the respective space, $X$ and $Y$ represent the input and output feature spaces. 
In this paper, we focus on the rotation equivariance in CNN, since rotation symmetry is prevalent in dynamic MR data. It has been demonstrated that the equivariance of the entire unrolled MRI reconstruction model can be achieved as long as the proximal operator satisfies rotation equivariance.\cite{ref3}. Therefore, it is desirable to model the proximal operator defined in section 2.1 with rotation-equivariant CNN. In most advanced dynamic MRI reconstruction CNNs, the 3D convolutions are replaced with (2+1)D convolutions consisting of 2D spatial and 1D temporal components \cite{ref17,ref20}. However, current equivariant CNNs can only be applied to 2D or spatial 3D convolutions \cite{ref24,ref25}, which have not been explored for 2D+t data. Directly applying 2D equivariant convolution methods to (2+1)D CNNs is unable to achieve equivariance, since the 1D temporal convolution layer will destroy the equivariance between 2D equivariant convolution layers, consequently destroying the global equivariance of the network. 

To preserve the equivariance of the network, a temporal-equivariant convolution layer is designed and applied to construct the SRE-CNN, which is shown in Fig. 2(b). The input, output and intermediate equivariant convolution layers have been well defined in previous works \cite{ref6,ref24}, so we only give a brief description. We define a rotation transformation group $S$ with $s$ elements. The rotation operators $A$ and $B$ are group elements of $S$, which are used to rotate the filters. 

\begin{figure}[t]
\includegraphics[width=\textwidth]{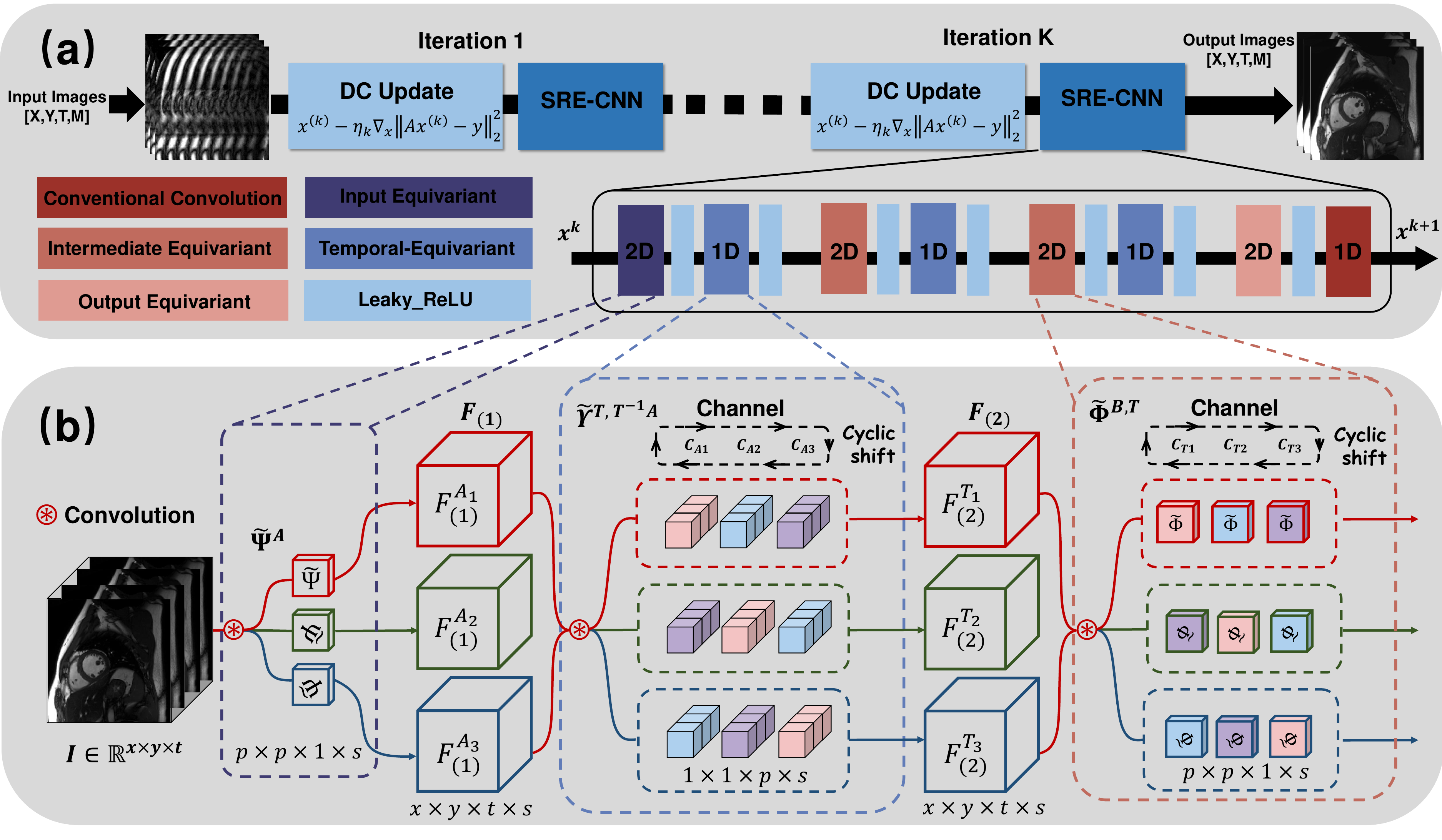}
\caption{(a) The unrolled reconstruction model with the PGD method. The layers in SRE-CNN are signed by different colors. (b) Illustration of an example network constructed by the proposed SRE-CNN, where we set the rotation transformation group $S ( A_i,B_i,\in S)$ as ${2\pi i}/3$ rotations, $i=1,2,3$. The proposed temporal equivariant layer, input and intermediate equivariant layer are marked by dashed lines. $p$ denotes the size of filter. $F$ denotes feature maps layer.}\label{fig2}
\end{figure}

In input equivariant layer, the 2D filters learn spatial patterns in different frames of input 2D+t data. Each set of adjacent filters share a same weight and rotated by the action in group $S$ to ensure the feature robustness to input image rotation. Thus, similar features with different orientations can be responded. $\tilde{\Psi }$ denotes the discrete convolution filters, $\ast$ is regular discrete 2D convolution, $\star$ is group convolution \cite{ref6}. For any $A\in S$, the convolution in input equivariant layer can be expressed as: $\left ( \tilde{\Psi }\star I\right )^{A} = {\tilde{\Psi }}^A\ast I$. The feature maps after the input equivariant layer denote $F_{(1)}$. Each color of the feature map in $F_{(1)}$ is corresponding to a different rotation element $A$ in group $S$.

The proposed temporal-equivariant layer can be formulated as:
\begin{equation}
(\tilde{\Upsilon} \star F)^T = \sum_{A\in S}^{} {\tilde{\Upsilon}^{T,T^{-1}A}*F^A}
\end{equation}
where $\tilde{\Upsilon}$ denotes the filters. As shown in Fig. 2(b), the channels of each 1D filter are extended to match the additional channels in $F_{(1)}$. Specifically, the channel $C_{A1}$ of the filter should be convolved with the feature map $F_{(1)}^{A_1} $. To preserve the equivariance, the channel orders of the filters in a same group are cyclically shifted to match each rotation element in group $A$. The same color of channels in each filter represents the same weighting. As 1D Temporal convolution is not required to rotate in $x\times y$ spatial dimension, $T$ are not rotation actions but still have $s$ elements to match $A$ and $B$. 

In a intermediate equivariant layer, $\tilde{\Phi}$ denotes the convolution for intermediate layer. For any $B\in S$, the convolutions in this layer can be expressed as: $(\tilde{\Phi} \star F)^B = \sum_{T\in M}^{} {\tilde{\Phi}^{B,T}*F^{BT}}$. The channels of the filters still need to be cyclically shifted. Due to it is a 2D convolution layer, the spatial dimension of filter should be rotated by the action element in group $B$ to maintain the relative orientation with the original input. Output equivariant layer is to reduce the additional channels extended by rotation group. 

Actually, the proposed temporal-equivariant convolution layer can maintain the rotation equivariance between any kind of 2D equivariant layer, thereby preserving the global equivariance of (2+1)D CNN.

\subsection{Filter Parametrization}
2D Filter parametrization technique has been widely used in equivariant CNNs to achieve an arbitrary angular resolution w.r.t. the sampled filter orientations without suffering interpolation artifacts from filter rotation \cite{ref18}. However, the current filter parametrization methods still remain the problem of low representation accuracy, which is not sufficient for reconstruction of image details.\cite{ref19,ref23}. To address this issue, we present a high-precision filter parametrization method based on 1D and 2D Fourier series expansion, since Fourier series expansion is equivariant to inverse discrete Fourier transform, which is with zero representation error \cite{ref25}. Specifically, we define the filters as the linear combination of a set of basis functions and learn the combination coefficients. For 1D temporal-equivariant convolutions, we adopt 1D Fourier series expansion as the basis functions to construct filters, which can be expressed as:
\begin{equation}
\tilde{\phi}\left ( x \right ) = \sum_{k=1}^{p-1} \left ( a_{k} \tilde{\varphi}_{c}^{k} \left ( x \right ) + b_{k} \tilde{\varphi}_{s}^{k} \left ( x \right )\right ) 
\end{equation}
where $\tilde{\phi}$ is the filter, $\tilde{\varphi}_{}^{k}$ are basis functions, i.e. 1D Fourier series expansion, $p$ denotes the filter size, $a$ and $b$ are learned coefficients.
Nonetheless, the aliasing effect in 2D Fourier series expansion for the rotated cases is significant. Followed by Xie's work \cite{ref25}, we replace high-frequency bases with the mirror functions of low-frequency bases to alleviate this problem and apply it to construct filter in 2D equivariant convolutions, which can be expressed as:
\begin{equation}
\varphi_{c}^{kl} \left(x\right) =\cos \left ( \frac{2\pi}{ph} \left [k-\left \lfloor \frac{p}{2} \right \rfloor,l- \left \lfloor \frac{p}{2} \right \rfloor \right ] \cdot \begin{bmatrix}x_1
 \\ x_2 \end{bmatrix} \right ) ,
\end{equation}
$h$ is the mesh size of images. With the help of such high-accuracy filter parametrization,  SRE-CNN can effectively leverage the rotation symmetry of local features. 

\section{Data and Experiments}

We use a set of dynamic cardiac cine data to evaluate our approach. Informed consent was obtained from the imaging subjects in compliance with the Institutional Review Board (IRB) policy. The fully sampled cardiac cine data were collected from 29 healthy volunteers on a 3T scanner (MAGNETOM Trio, Siemens Healthcare) with 20-channel receiver coil arrays. For each subject, 10 to 13 short-axis slices were imaged with the retrospective electrocardiogram (ECG)-gated segmented bSSFP sequence during breath-hold. The following sequence parameters were used: FOV = 330$\times$330 mm, acquisition matrix = 256$\times$256, slice thickness = 6 mm, TE/TR = 1.5/3.0 ms. The acquired temporal resolution was 40.0 ms and each data has 25 phases that cover the entire cardiac cycle. We randomly selected 25 volunteers for training and the rest for testing. Data augmentation using rigid transformation-shearing was applied. Finally, 800 2D-t multi-coil cardiac MR data of size 192$\times$192$\times$18 ($x\times y \times t$) are used for training, 30 for validation, and 118 for testing. 

\begin{table}[t]
\caption{Quantitative analysis of reconstruction for accelerated cardiac cine MRI (R=12, 16 and 20) using the proposed SRE-CNN and competing methods. The mean value with standard deviations are shown. The best results are marked in bold.}\label{tab2}
\centering
\begin{tabular}{l|l|l|l|l}
\hline
Acc R ~~ &  Methods & PSNR $\uparrow$ & SSIM $\uparrow$ & Time (s)\\
\hline
12 &  L+S                   & 37.5190(2.3518) & 0.9331(0.0347) & 94.56 s\\
   &  MoDL                  & 42.4620(1.6709) & 0.9628(0.0099) & 0.52 s\\
   & R2plus1D               & 44.3743(1.8894) & 0.9754(0.0083) & 0.38 s \\
   & Proposed SRE-CNN ~~~~ & \bfseries 45.9801(2.0681) ~ & \bfseries 0.9900(0.0063) ~ & 0.93 s\\
\hline
16 &  L+S                   & 33.9077(2.0584) & 0.8720(0.0493) & 94.56 s\\
   &  MoDL                  & 41.1211(1.8775) & 0.9607(0.0118) & 0.52 s\\
   & R2plus1D               & 41.9028(1.8708) & 0.9644(0.0116) & 0.38 s \\
   & Proposed SRE-CNN ~ & \bfseries 43.7158(2.0413) & \bfseries 0.9828(0.0083) & 0.93 s\\
\hline
20 &  L+S                   & 30.0754(1.8259) & 0.7878(0.0536) & 94.56 s\\
   &  MoDL                  & 39.2965(1.5566) & 0.9448(0.0134) & 0.52 s\\
   & R2plus1D               & 40.2366(1.8249) & 0.9536(0.0137) & 0.38 s \\
   & Proposed SRE-CNN ~ & \bfseries 42.1974(2.0528) & \bfseries 0.9761(0.0102) & 0.93 s\\
\hline

\end{tabular}   
\end{table}

The number of unrolled iterations is set to be 10 empirically, four layers with the filter number of 46-46-46-2 in each iteration are used. The ADAM optimizer was employed, and the learning rate was set to 0.001 with an exponential decay rate of 0.95. The loss function is constructed as $l_1$  differences in pixel-wise sense. The training and testing were on a workstation equipped with a Nvidia Tesla V100 Graphics Processing Unit (GPU). The fully sampled data were retrospectively under-sampled with VISTA \cite{ref2} of R=12, 16 and 20 acceleration to generate the pair of training samples. 

We compared our proposed approach with SOTA dynamic MR reconstruction methods, including a compressed sensing (CS) method L+S \cite{ref15}, a DL-based method dynamic MoDL \cite{ref1} and DL-ESPIRiT (named as R2plus1D) \cite{ref17}. All the DL methods were designed with a parameter size of around 330k to give fair comparisons. Peak signal to noise ratio (PSNR) and structural similarity index (SSIM) \cite{ref22} were employed to evaluate the reconstruction.

\section{Results and Discussion}

\begin{figure}
\includegraphics[width=\textwidth]{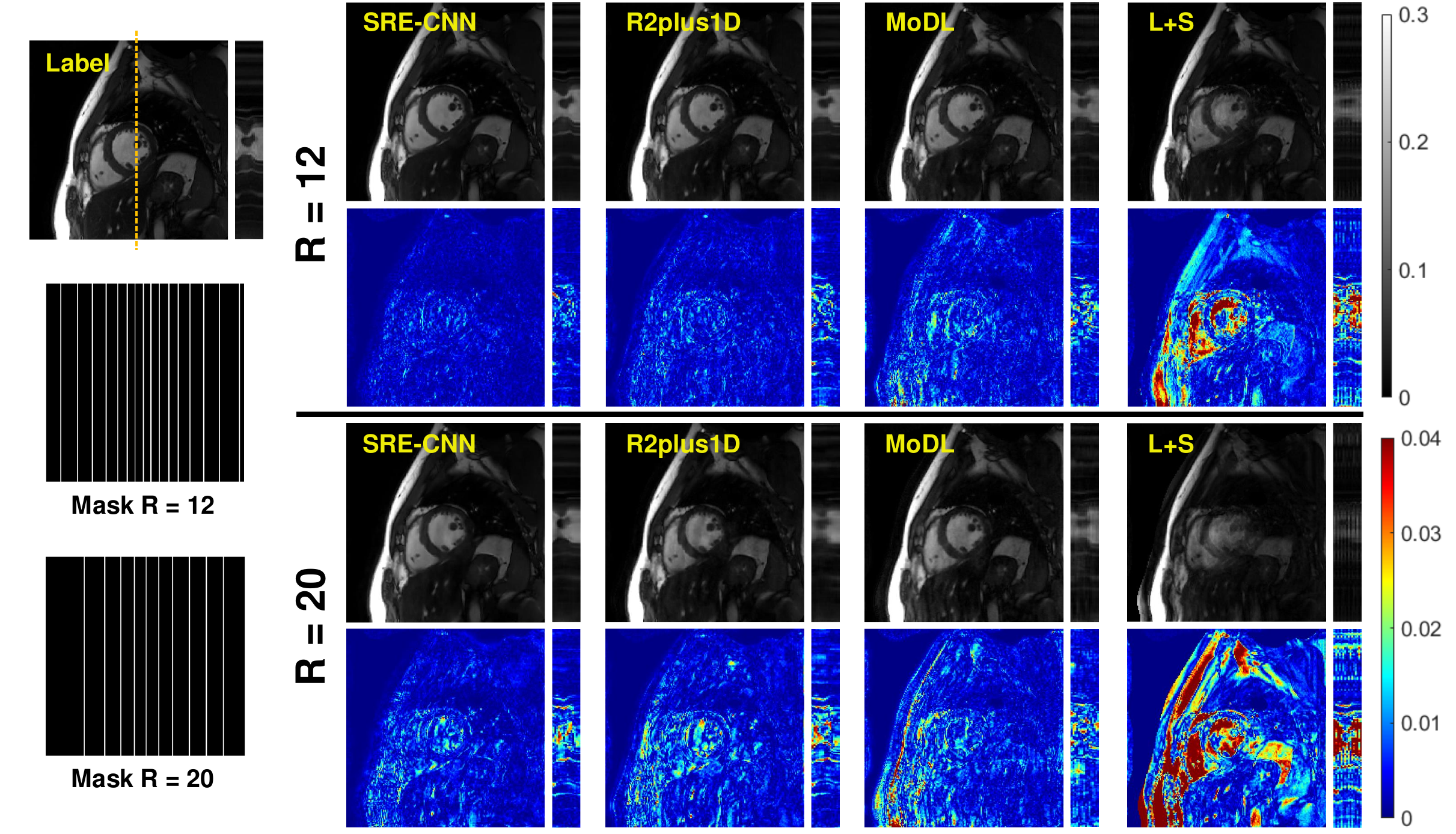}
\caption{Reconstructions in the spatial domain and along the temporal dimension (x-t image) under acceleration (R=12 and 20) and the corresponding error maps.} \label{fig3}
\end{figure}

The quantitative results in Table 1 show consistent superior performance of the proposed method across every single undersampling factor compared to all other baseline methods. At extremely high acceleration factors (16X, 20X), the proposed SRE-CNN still gives superior reconstructions by a large margin ($\sim$ 2dB in PSNR). DL methods could achieve much faster reconstruction than CS methods since only one forward process is required in the test mode. The proposed SRE-CNN achieves better performance in higher accelerated factor, which indicates the prior knowledge of rotation symmetry is useful in highly undersampled MRI reconstruction. Although the proposed SRE-CNN has a little longer reconstruction time than other DL methods due to the equivariant convolutions, its reconstruction time is still less than 1s, which is acceptable.

The qualitative comparisons are shown in Fig. 3, where the reconstructed images in the spatial domain and temporal domain, as well as the corresponding error maps, were provided. From the figure, the proposed SRE-CNN can faithfully reconstruct the images with smaller errors and clearer anatomical details. Importantly, the edge of myocardium in reconstructed images, i.e. the regions with more rotated variants of the similar anatomical structure were evidently better reconstructed. It indicates that SRE-CNN can effectively exploit the rotation symmetry present in images and significantly improve the quality of details reconstruction. From x-t results, it can be observed that our proposed approach can still recover the dynamic information very well from extremely high undersampled data (R = 20), where the conventional CS approaches fail and the competing deep learning approaches cannot remove artifacts well.

It is noted that all of our proposed modules can be universally adopted by most CNN-based model. Our filter parametrization method can be applied to improve the representation accuracy of any parameterized filter in 1D and 2D convolutions. The SRE-CNN module with the designed temporal-equivariant convolution layer can be used in most CNNs to effectively exploit the rotation symmetry in (2D+t) data, such as video tasks.

\section{Conclusion}
In this study, we have proposed a Spatiotemporal Rotation-Equivariant CNN (SRE-CNN) with a high-accuracy filter parametrization strategy to take full advantage of the rotation symmetry in dynamic MRI reconstruction. The improved reconstruction performance has demonstrated that our method can effectively utilize the rotation symmetry along the temporal dimension and the rotation symmetry of local features within image frames to enhance the reconstruction of detailed features in image. In our future work, we will further extend the employed rotation temporal equivariant CNN to a larger group of transformations, such as the composition of scale and rotation.

\begin{credits}
\subsubsection{\ackname} This study was funded by the National Natural Science Foundation of China (62106252, 62331028, 62206273 and 62125111), Key Laboratory for Magnetic Resonance and Multimodality Imaging of Guangdong Province (2023B1212060052). 

\subsubsection{\discintname}
The authors have no competing interests to declare that are relevant to the content of this article.
\end{credits}


\end{document}